\documentclass[a4paper]{article}

\usepackage[utf8]{inputenc}
\usepackage{erk}
\usepackage{times}
\usepackage{graphicx}
\usepackage{amsmath}
\usepackage{amssymb}
\usepackage[top=22.5mm, bottom=22.5mm, left=22.5mm, right=22.5mm]{geometry}

\usepackage[slovene,english]{babel}

\def\footnotemark{}

\begin{document}
\title{Solution space of optimal heat pump schedules}

\author{David Kralji\'{c}$^{1}$, Miha Troha$^{1}$} 

\affiliation{$^{1}$COMCOM d.o.o Idrija, Lapajnetova 29, Idrija, Slovenia}

\email{E-pošta: david.kraljic@comcom.si}

\maketitle

\begin{abstract}{Abstract}
We study the space of optimal schedules for a heat pump with thermal energy storage used in heating a residential building. We model the heating system as a Mixed Integer Linear Program with the objective to minimise the cost of heating. We generate a large number of realistic daily heat demands and calculate the optimal schedule for the heat pump. In addition to cost savings stemming from optimal running, we find that the space of optimal schedules is large in practice, even for the simplest model of the heating system we use, and that the optimal schedules are difficult to reproduce with statistical models. These findings strengthen the case for the use of mathematical optimisation in real-life applications.
\end{abstract}

\selectlanguage{english}

\section{Introduction}
Mathematical optimisation is a generic framework for formulating and solving problems where some goal (or objective) needs to be optimised (maximised or minimised) given some constraints. It has been applied successfully across many domains such transport and logistics\cite{app_transport}, power plant dispatch and electric grid balancing \cite{energy}, economics \cite{economics}, machine learning \cite{ml}, biology \cite{biology} and many others.

With the advent of smart buildings and smart devices\footnote{Examples include air conditioning units, heat pumps, charging electric cars.} which enable users to actively control them, the problem arises of how to do that optimally. Mathematical optimisation framework is well suited to tackle this problem. The goal of a smart device owner is to minimise utility costs (or equivalently maximise profits) subject to the physical characteristics\footnote{Examples include size, losses, speed of charging or discharging etc.} of a smart device.

Usually, in practice, smart devices are not actively controlled by mathematical optimisation algorithms, but instead they follow simple rules or some predefined daily schedules. The question appears whether modelling the devices and surrounding processes to produce their optimal schedules adds any value in comparison to schedules or logic fixed in advance. The aim of this contribution is to study this question in a simplified but realistic heating system for an actual residential building. We focus on the properties of the solution space for optimal running of the heating system given the realistic demand for heat for the building. 

In Section\,\ref{sec:setup} we describe the devices and other data, in Section\,\ref{sec:opti} we discuss the mathematical and statistical modelling used, in Section\,\ref{sec:results} we show the results and discuss them.

\section{Set-up, Data and Benchmark}
\label{sec:setup}
The optimisation problem that we address in this contribution aims to minimise costs of operating a heat pump by considering physical constraints of the heat pump and the requirement that enough heat must be generated to satisfy heat demand of the building. Data for costs, physical constraints of the heap pump and demand of the building were determined based on realistic historical data of a residential building\footnote{Retirement home `Dom Sv. Lenart', Lenart, Slovenia}. 

\subsection{Costs}
The main cost of heating comes from the cost of electric energy which is split into two tariffs: lower cost tariff (MT) applicable between 10pm and 6am and a higher cost tariff (VT) applicable between 6am and 10pm. We assume that VT stands at $1.5$ of MT, which is a realistic split given the current electricity prices.

\subsection{Physical constraints}
The heat pump is capable of providing roughly 160\,kW of heat at a coefficient of performance\footnote{The coefficient of performance determines the heat generated by one unit of electrical power.} $\sim1.6$. The heat storage is capable of storing up to 200\,kWh of energy.

\subsection{Demand}
\begin{figure}[htb]
    \begin{center}
        \includegraphics[width=\columnwidth]{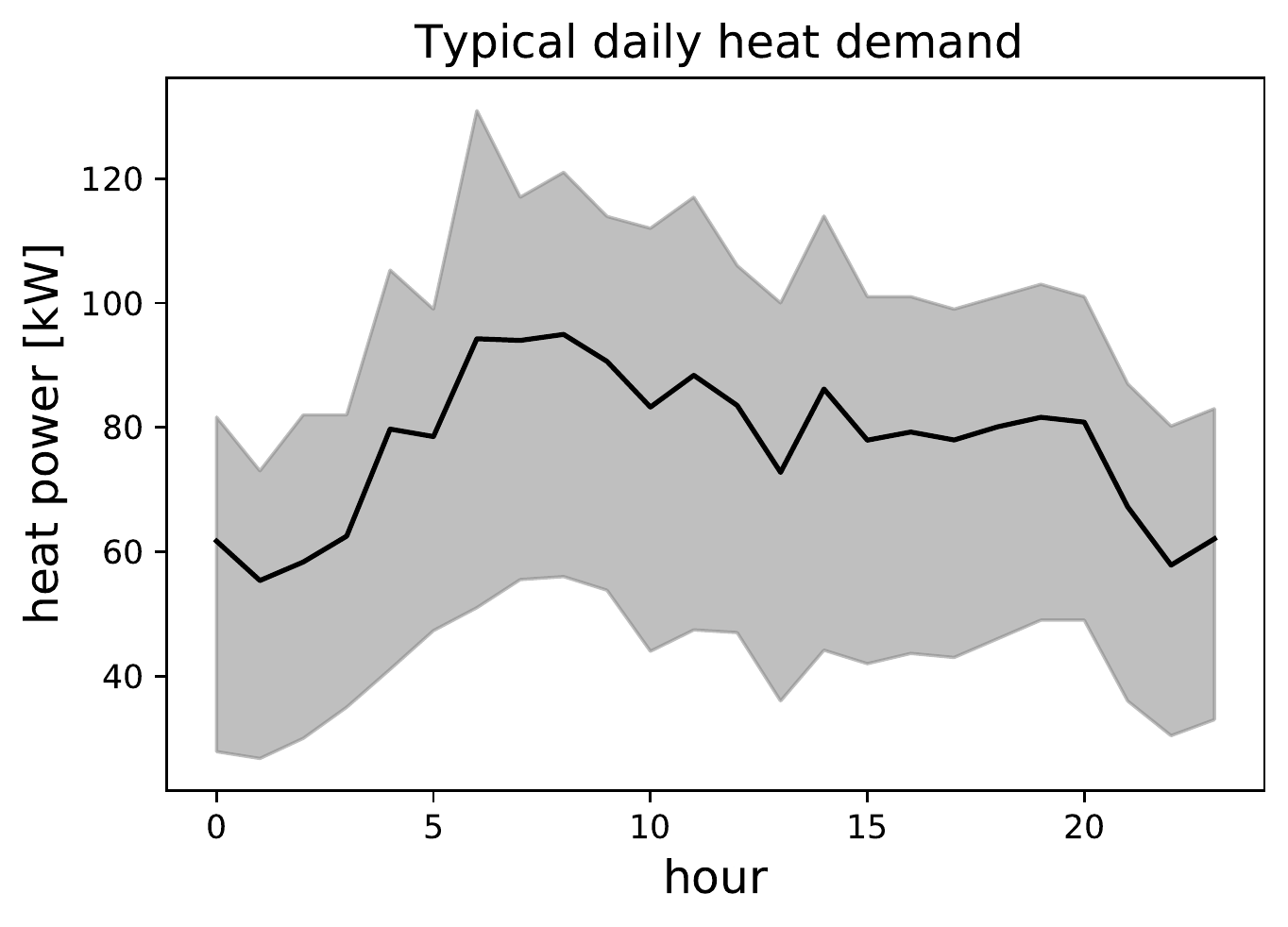}
        \caption{Typical daily heat demand. The shaded region around the mean line corresponds to area between the first and third quantile.} \label{fig:typical_heat_demand}
    \end{center}
\end{figure}

We use real heat demand data which is measured by a calorimeter at the exit of the heating system. A typical daily demand profile based on 480 days during three years of heating season (mid October to April) is depicted in Fig.\,\ref{fig:typical_heat_demand}. 

\subsection{Benchmark}
\begin{figure}[htb]
    \begin{center}
        \includegraphics[width=\columnwidth]{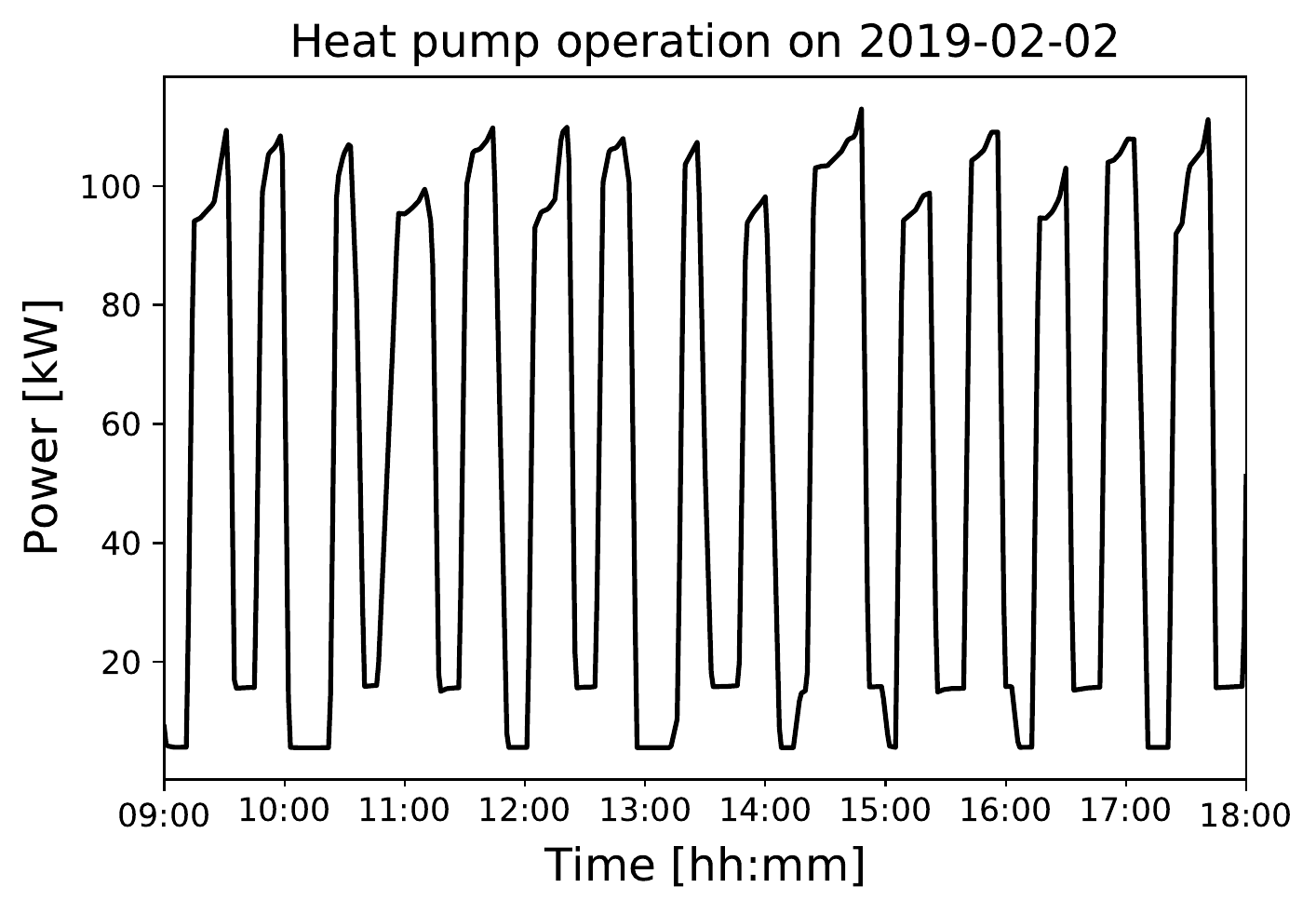}
        \caption{Typical operation of the heat pump.} \label{fig:typical_heat_pump}
    \end{center}
\end{figure}

The heat pump operates in the `on/off' fashion, that is, either it runs at maximum power or is idle at low power. Schedules produced by the mathematical optimisation algorithms are to be benchmarked against actual historical schedules of the heat pump. For an example of a historical schedule, as captured by a power meter located at the heat pump, see Fig.\,\ref{fig:typical_heat_pump}. 

\section{Methods}
\label{sec:opti}
\subsection{Optimal schedule of the heat pump}
\label{sec:opti_sch}
We obtain the optimal daily schedule for the heat pump by building an optimisation model for the heating set-up with the objective to minimise the cost of operation. We follow the modelling of \cite{heat_pump_1}, \cite{heat_pump_2}.

Our aim is to evaluate the variability of optimal solutions given the heat demand. Therefore, we simplify a realistic description as far possible without losing the crucial elements of the set-up - that is the variation in schedules is only determined by the variations in heat demand and not in the variation of the modelling parameters. Thus, we neglect the variable, fixed, and cycling costs of the heat pump as well as heat losses in the storage. We also assume the coefficient of performance is constant throughout the day (even though it varies with outside temperature). We also model the heat pump as only being on at maximum power or off. Finally, we chose hourly granulation for all the variables and parameters in our model. This means that the solution space for the heat pump can be encoded as a sequence of zeros and ones of length 24, corresponding to heat pump either running or not in a specific hour. The size of the solution space is $2^{24}$.

The optimisation problem we solve reads as follows:
\begin{align}
    \min_{P_{t}, \overline{P}_{t}, S_{t}} &\qquad \sum_{t\in T} P_{t}^{\text{HP}} \cdot \pi_{t}^{\text{elec}} \label{eq:opti_obj}\\
    \text{s. t. } &\qquad S_{t+1} = S_{t} + \text{cop}_{t} \cdot P_{t}^{\text{HP}} - \text{hd}_{t} \; t \in \overline{T} \label{eq:opti_storage}\\
     &\qquad P_{t}^{\text{HP}} = p^{\max} \cdot \overline{P}_{t}^{\text{HP}} \; t \in T \label{eq:opti_onoff}\\
     &\qquad s^{\min} \leq S_{t} \leq s^{\max} \; t \in T \label{eq:opti_storage_minmax}\\
     &\qquad P_{t}, S_{t} \in \mathbb{R} \; t \in T\\
     &\qquad \overline{P}_{t} \in \mathbb{B} \; t \in T
\end{align}
where $P_{t}^{\text{HP}}$ is the heat pump power, $\pi_{t}^{\text{elec}}$ is the electricity price, $S_{t}$ is the heat energy stored, $\text{cop}_{t}$ is the coefficient of performance, $\text{hd}_{t}$ is the building heat demand, $p^{\max}$, $s^{\max}$, $s^{\min}$ are the power and energy limits, and $\overline{P}_{t}^{\text{HP}}$ is the binary decision variable denoting whether the heat pump is on or off. Time sets $T$ and $\overline{T}$ are defined as $\{0, 1, ..., 23\}$ and $\{0, 1, ..., 22\}$.

The objective in Eq.\,\ref{eq:opti_obj} is the cash paid for the electricity consumed to heat the building, which we minimise\footnote{We add to electricity prices a very small random increase with each hour of the day, thus ensuring that the optimal objective has a unique solution that prefers taking decisions earlier in the day.}. Eq.\,\ref{eq:opti_storage} describes the energy balance for the heat storage. The heat pump is constrained to be either off or on at max power whcih is enforced by Eq.\,\ref{eq:opti_onoff}. The storage limits are set by Eq.\,\ref{eq:opti_storage_minmax}.

We use the following parameters: $p^{\max} = 100\text{\,kW}$, $s^{\max} = 200\text{\,kWh}$, $s^{\min} = 0\text{\,kWh}$, $\text{cop}_{t} = 1.6$. For all runs we set the initially stored heat, $S_0 = 100\text{\,kWh}$.

\subsection{Daily heat demand profiles}
\label{sec:demand_profiles}
We are interested in running the optimisation problem from Sec.\,\ref{sec:opti_sch} a large number of times to explore as much of solution space as computationally possible. For a 24 hour granulation of the daily optimal schedule of an `on/off' heat pump the total number of potential solutions is $2^{24} \approx 1.7\cdot 10^7$. Since we only have the actual data for 480 days during the heating season, we need to generate synthetic but realistic heat demand profiles. 

\begin{figure}[htb]
    \begin{center}
        \includegraphics[width=\columnwidth]{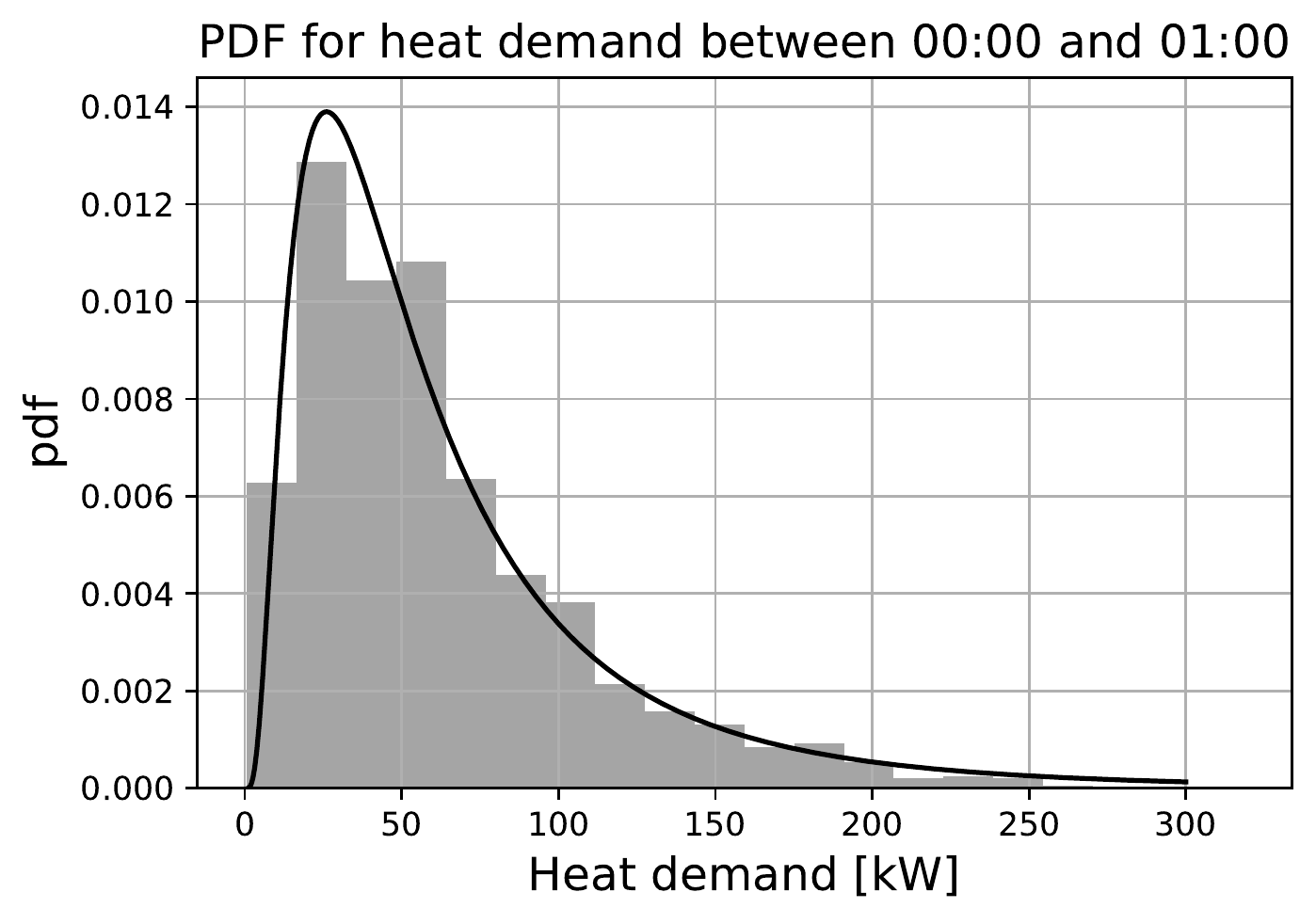}
        \caption{Shaded is the empirical pdf for the heat demand between hour 00:00 and 01:00. The line is our best fit lognormal model.} \label{fig:pdf_fit}
    \end{center}
\end{figure}

Heat demand for each hour of the day is a random variable with a mean and a variance in the simplest terms. Because the demand is bounded from below with 0, a simple normal distribution cannot be used to model it. We chose a lognormal distribution\footnote{logarithm of a random variable is distributed normally} which is bounded by 0 and is also only described by a mean and variance. We illustrate our modelling for a specific hour in Fig.\,\ref{fig:pdf_fit}.

\begin{figure}[htb]
    \begin{center}
        \includegraphics[width=0.8\columnwidth]{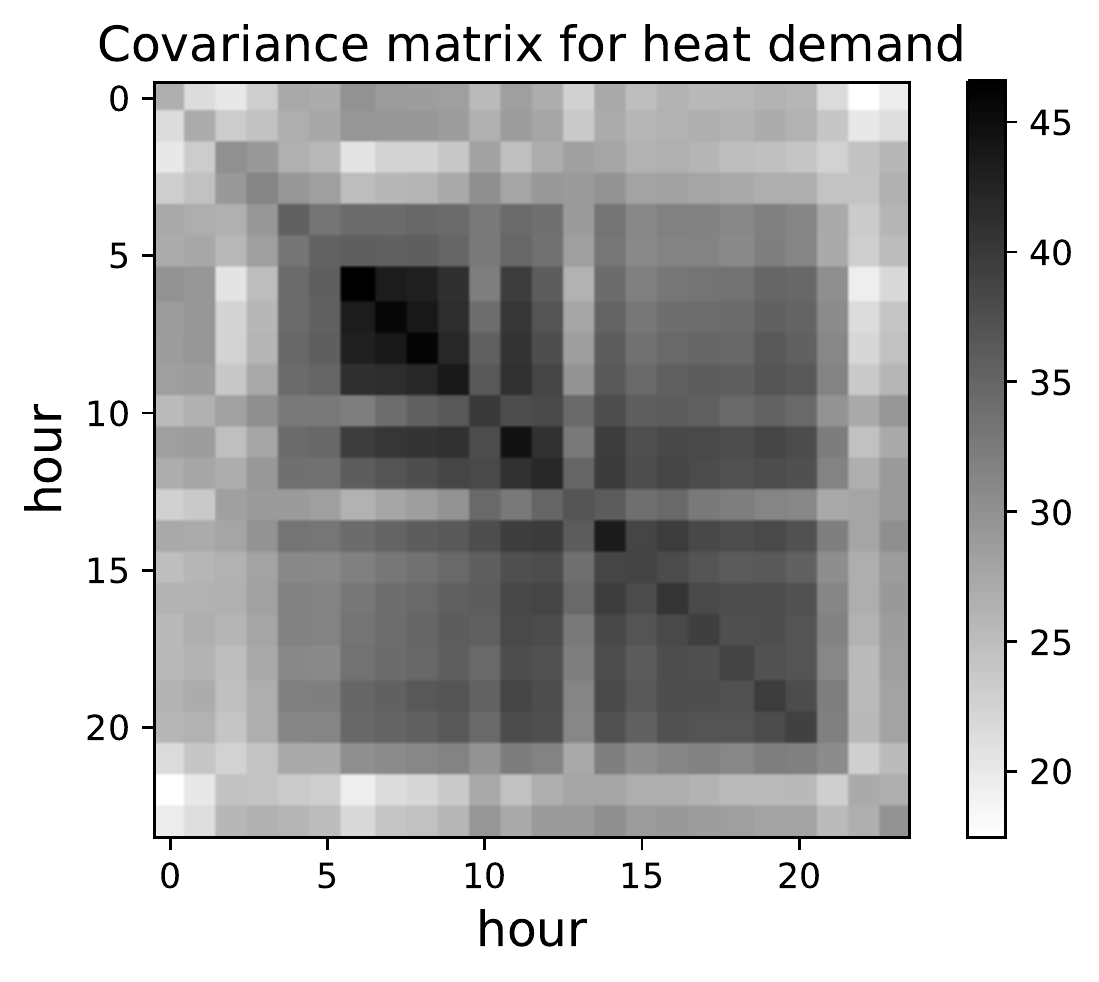}
        \caption{Empirical covariance matrix for the hourly heat demand.} \label{fig:covariance}
    \end{center}
\end{figure}

Because the heat demand from hour to hour within a single day is not independent, we need to also model these hourly cross-correlations. The size of these is illustrated by the covariance matrix for the hourly demand in Fig.\,\ref{fig:covariance}.

Our final model for the daily demand is the lognormal distribution derived from empirical mean (Fig.\,\ref{fig:typical_heat_demand}) and the covariance matrix (Fig.\,\ref{fig:covariance}) of the actual data.

\section{Results \& Discussion}
\label{sec:results}
We have generated $100\,000$ daily heat demand profile samples and solved the optimisation problem (Sec.\,\ref{sec:opti_sch}) the same number of times, obtaining $100\,000$ optimal solutions.

\subsection{Size of solution set}
To determine if our solution set is localised in the space of all possible solutions, we compare it to the baseline case -- the set where all $2^{24}$ solutions for the heat pump schedule are possible and equally probable. For such a uniformly distributed set a single solution appears on average $\lambda = n/2^{24}$ times, where $n=100\,000$ is the number of generated solutions. Then it follows that the \emph{number} of times a solution appears is Poisson distributed with mean $\lambda$.

\begin{figure}[htb]
    \begin{center}
        \includegraphics[width=\columnwidth]{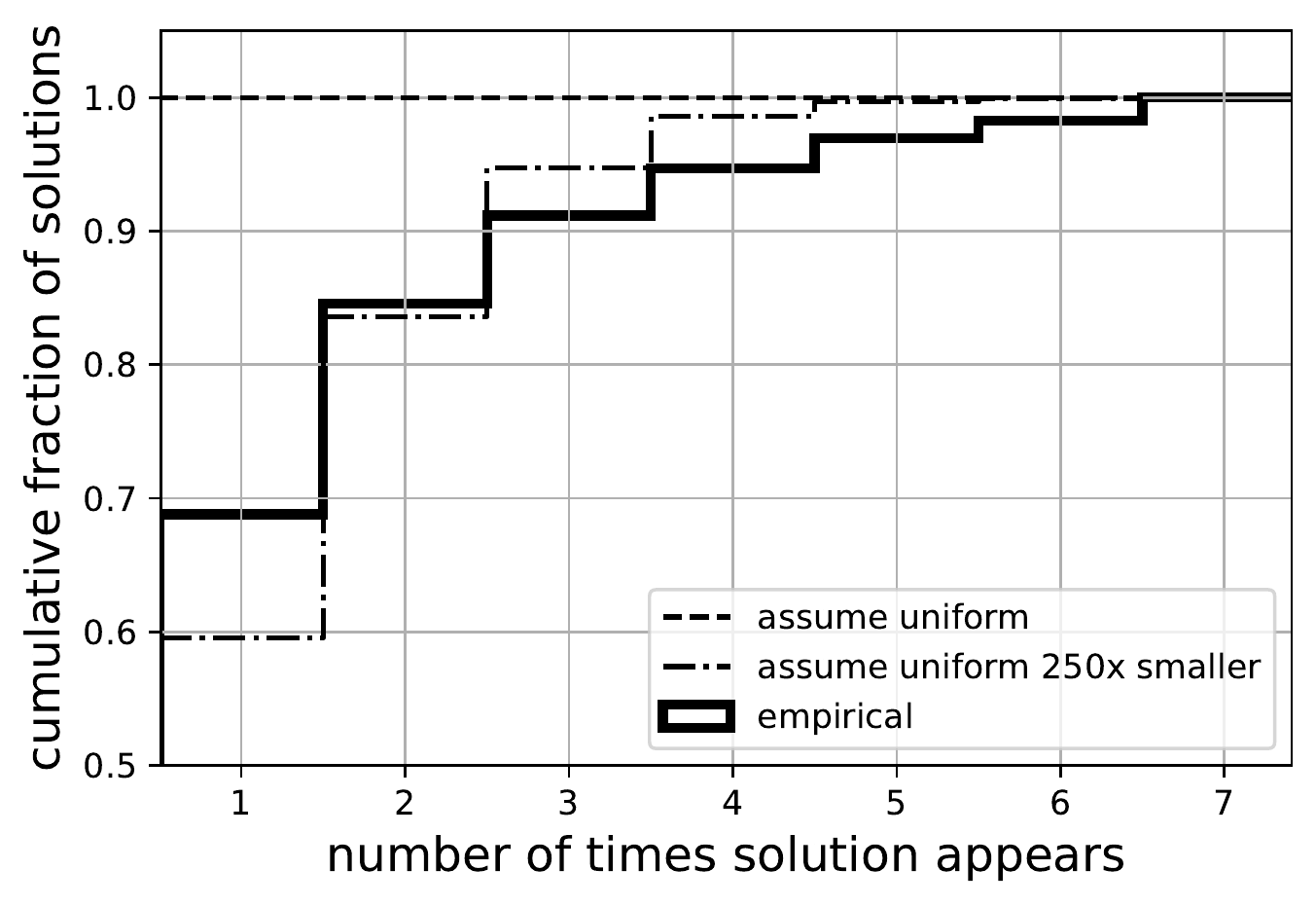}
        \caption{Empirical multiplicities of our solution set compared to expectations if solutions were uniformly distributed in the space of all possible solutions.} \label{fig:multiplicity}
    \end{center}
\end{figure}

In Fig.\,\ref{fig:multiplicity} we plot the fraction of solutions we found against the multiplicity of the solutions. That is, we plot the fraction of solutions that appear at most once, twice, three times etc., in our solution set. We also plot the expectations stemming from the uniform distribution of solutions, as explained above. 

We see (Fig.\,\ref{fig:multiplicity}) that the set of optimal schedules we obtain has solutions appearing more often and at higher multiplicities than expected if the baseline case would be true. Note, that the fraction of solutions that appear only once in the baseline case is practically all of them (0.9999823). This suggests that the solution space to our optimisation problem is localised and not completely random. This is expected, as the variability in the heat demand is not completely random, but constrained.

However, the fraction of solutions that appear once, twice etc., is still small. For example, there are no solutions that would appear more than 7 times in our set of $100\,000$, suggesting that different demand profiles lead to very different optimal solutions. 

If we assume the solution set to our optimisation problem is uniformly distributed (but smaller) then by fitting the Poisson distribution to our empirical data in Fig.\,\ref{fig:multiplicity} we deduce that the solution space is roughly 250 times smaller than $2^{24}$ ($\approx7\cdot10^{4}$). This means a single solution to our optimisation problem is expected to repeat roughly every $7\cdot10^{4}$ daily runs or 180 years.

\subsection{Complexity of solution set}

We measure how easy it is to describe the solutions to our heat pump optimisation problem by training machine learning models to predict optimal heat pump schedules given the heat demand and evaluating their performance\footnote{Machine learning models have been used extensively in predicting solutions to optimisation problems \cite{mlmip1, mlmip2, mlmip3, mlmip4}.}. 

The generated dataset of $100\,000$ heat demand profiles forms the input set with the corresponding optimal solutions being the output set. We split the dataset into train, validation, and test subsets. We selected tree-based (Decision tree, Random Forest) models and linear models (Logistic regression)\footnote{These models are selected as they can be expressed easily in code and thus can be deployed to control the heat pump}. All model parameters were fine-tuned on the validation set, with the final performance evaluated on the test set. We also evaluate a trivial model which produces random schedules.

We measure how well machine learning models predict the optimal solution by counting the average number of hours that differ between true and predicted schedules. In other words, we count the minimum number of errors that transform the true optimal solution to the one predicted by a machine learning model (known as the Hamming distance\cite{hamming} in coding theory). We summarise the results in Table\,\ref{tab:ml}.
\begin{table}[h]
\caption{Performance of machine learning models predicting optimal solutions.} \label{tab:ml}
\smallskip
\begin{center}
\begin{tabular}{ | r | c | c | }
\hline  
  \textbf{Model} & \textbf{Best param.} & \textbf{Hamming}\\ 
                 &                      & \textbf{error [hour]}\\ 
\hline
  Trivial model & - & 10.1\\
 \hline
  Decision Tree & max\_depth = 10 & 5.94\\
\hline
  Random Forest & max\_depth = 10 & 5.72\\
                & num\_trees = 400 & \\
\hline
  Log. Regression & - &  6.36 \\
\hline  
\end{tabular}
\end{center}
\end{table}

Solving the optimisation problem in Sec.\,\ref{sec:opti_sch} performs a mapping from the daily heat demand profile to the optimal schedule. Mimicking this mapping with either linear or tree-based models does not work well since on average the predicted optimal solutions are still wrong on about 6 hours of the day.

\subsection{Cost savings}
\begin{figure}[htb]
    \begin{center}
        \includegraphics[width=\columnwidth]{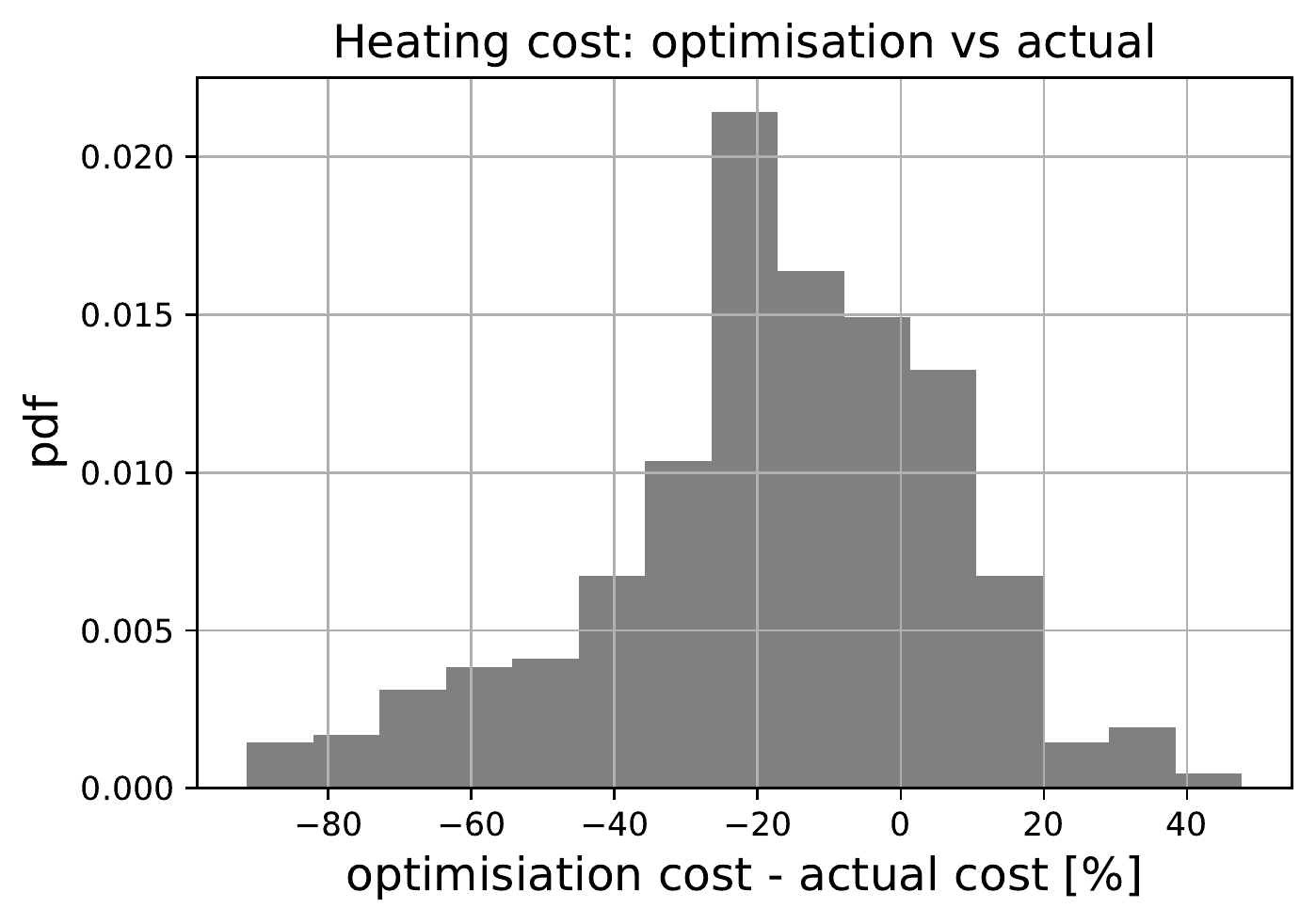}
        \caption{Percentage difference between the costs of schedules produced by our simple optimisation model and actual costs.} \label{fig:costs}
    \end{center}
\end{figure}

Currently (i.e. in our benchmark data), the heat pump operates in short bursts of 15-30minutes (e.g. Fig.\,\ref{fig:typical_heat_pump}), and follows the real-time demand for heat almost exactly without utilising the large heat storage available. We evaluate the cost of this current strategy on the 480 days of heating seasons using the actual data. We run the simplified optimisation problem from Sec.\,\ref{sec:opti_sch} for the same days and plot the differences in Fig.\,\ref{fig:costs}

We note that even the simple modelling of the heating system we use manages to save on average 15\% in heating costs. We expect that formulating the optimisation problem with higher granularity (sub-hourly), as well as using time-varying COP, would make the gains even bigger, as the problem would be less constrained.

\section{Conclusion}
We studied the set of optimal solutions for a simplified heating system consisting of a heat pump\&storage satisfying realistic heat demand profiles. The set-up and data are based on an actual residential building. We find that the solution space for optimal heat pump schedules is smaller than the maximum solution space, but still large enough to justify using the optimisation framework in practice. We also show that the optimal schedules, once they are known, are not easy to predict using statistical models. Finally, we show that even simplistic modelling of the optimisation problem can lead to significant cost savings.

\section*{Acknowledgement}
This contribution was done as part of the `BETAi' project supported by public tender `DEMO PILOTI II 2018', that is partly financially supported by European Union from European Regional Development Fund and Ministry of Economic Development and Technology.

We thank Martin Turk for data preparation and processing and Gasper Zadnik for implementing the optimisation problem in code.

\end{document}